\documentclass[a4paper,fleqn]{cas-dc}
\usepackage[compress,numbers]{natbib}
\usepackage[scaled]{helvet}
\usepackage[toc,page]{appendix}
\usepackage{mathtools}
\usepackage{booktabs}
\usepackage[english]{babel}
\usepackage{amsmath,amssymb,amsbsy,amstext, amsthm, simplewick, amsfonts}
\usepackage{hyperref}
\usepackage{graphicx}
\usepackage[small]{caption}
\usepackage{siunitx}
\usepackage{upgreek}
\usepackage{framed}
\usepackage{wrapfig}
\usepackage{multirow}
\usepackage{bbm}
\usepackage[utf8x]{inputenc}
\usepackage{selinput}
\usepackage{bm}
\usepackage{float}
\usepackage{geometry}
\usepackage{yfonts}
\usepackage{caption}
\usepackage{subcaption}
\usepackage{sidecap}
\usepackage{longtable}
\usepackage{anyfontsize}
\usepackage{dsfont}
\usepackage{tikz}
\usepackage{relsize}
\usepackage{tcolorbox}
\usetikzlibrary{matrix}
\usepackage{bigstrut}
\usepackage{fnpct}
\usetikzlibrary{arrows.meta}
\usetikzlibrary{decorations.pathreplacing,calligraphy}
\usepackage{tensor}
\usepackage{mleftright}
\usepackage{slashed}

\usepackage{soul}

\newcommand{\dint}{\text{d}}

\begin{document}
\let\WriteBookmarks\relax
\def\floatpagepagefraction{1}
\def\textpagefraction{.001}

\shorttitle{Dark Walker} 

\shortauthors{C. Yang} 

\title [mode = title]{Dark Walker in the Early Universe: A Strongly Coupled Sector Model}  

\author[1]{Chen Yang}[orcid=0000-0001-7127-0970]

\ead{chen.yang@sns.it}

\affiliation[1]{organization={Scuola Normale Superiore and INFN},
            addressline={Piazza dei Cavalieri 7}, 
            city={Pisa},
            postcode={56126}, 
            state={PI},
            country={Italy}}

\begin{abstract}
We explore the phenomenology of the ``Dark Walker''---an $\text{SU}(3)$ theory with eight flavors of massless fundamental fermions in the dark sector. 
During inflation, its walking dynamics generate primordial non-Gaussianities through the exchange of unparticles, while accounting for the current dark matter relic abundance if we consider freeze-in of Dark Walker coupled to the Standard Model through the Higgs portal. 
This provides a simple yet predictive example linking strongly coupled inflationary dynamics to present-day dark matter. 
\end{abstract}


\maketitle


\noindent{\bf Introduction.}---Assuming that the early universe underwent a phase of cosmic inflation, fields that weakly couple to the inflaton can play a crucial role in shaping the primordial non-Gaussianities---also referred to as cosmological correlators. 
One possible origin of scalar perturbations during inflation is the presence of composite particles, which arise from strongly coupled dynamics among more elementary constituents, similar to the mechanisms in quantum chromodynamics (QCD) and models of the composite Higgs. 
This scenario has been far less explored compared to the more conventional treatments of cosmological correlators involving fundamental particle exchanges. 
For recent developments in this direction, see, for example, \cite{Green:2013rd,Baumgart:2021ptt,Aoki:2023tjm,Chakraborty:2025myb,Pimentel:2025rds}. 

In this Letter, we consider a dark sector (DS) that interacts with the Standard Model (SM) through the Higgs portal and becomes strongly coupled during inflation, leaving imprints in the primordial fluctuations. 
After inflation, this DS contributes to today's relic abundance of dark matter (DM). 
We aim to explore the interplay between primordial non-Gaussianity and DM model building. 
As a simple illustrative example, we study what we call the ``Dark Walker''---a scenario that has the potential to account simultaneously for primordial non-Gaussianity $f_{\text{NL}}$ value at $O(1)$ and the observed DM relic abundance. 
\newline

\noindent{\bf Setup.}---To investigate a strongly coupled DS during inflation, it is useful to classify different scenarios according to the mass gap: for example, $\text{M}_{\text{Gap}}\ll H_{\text{inf}}$, $\text{M}_{\text{Gap}}\sim H_{\text{inf}}$ and $\text{M}_{\text{Gap}}\gg H_{\text{inf}}$. 
In this work, the Dark Walker model falls into the $\text{M}_{\text{Gap}}\ll H_{\text{inf}}$ regime, where the DS is well approximated by a conformal field theory. 
Such a conformal sector may contain operators with large anomalous dimensions, which are called \textit{unparticles}, as introduced in \cite{Georgi:2007ek,Grinstein:2008qk}, etc. 
An extended notion of unparticles with a mass gap has also been discussed in \cite{Strassler:2008bv}, but lies beyond the scope of this work. 
Gapless unparticles, however, cannot serve as DM, since their behavior resembles radiation rather than particles. 
Thus, the Dark Walker has to experience a phase transition, forming gapped states. 
During inflation, the model remains near the infrared fixed point (IRFP) long enough to leave imprints in the primordial non-Gaussianity. 
This behavior exemplifies a form of ``walking dynamics'', which gives rise to the name ``Dark Walker''. 

Let us consider a non-Abelian gauge theory in the DS, which is an $\text{SU}(N_c)_D$ theory with $N_f$ flavors of massless Dirac fermions. 
At the start of inflation, all modes are very blue-shifted and the effect of the de Sitter curvature in the $\beta$-function will be small, so we can approximately treat the background as flat space. 
The Lagrangian is 
\begin{align}
    \begin{aligned}
        \mathcal{L} &= -\frac{1}{4}F_{\mu\nu}^a F^{\mu\nu,a} + \bar{\psi}_ki\slashed{D}\psi_k, \ k=1,2,\dots,N_f \\
        D_\mu &= \partial_\mu + igA^a_\mu T_a. 
    \end{aligned}
    \label{def:UV_gauge_theory}
\end{align}
The $T_a$'s are the matrices of the adjoint representation of $\text{SU}(N_c)_D$ and $g$ is the gauge coupling. 

\begin{figure}[h]
    \centering
    \includegraphics[width=1\linewidth]{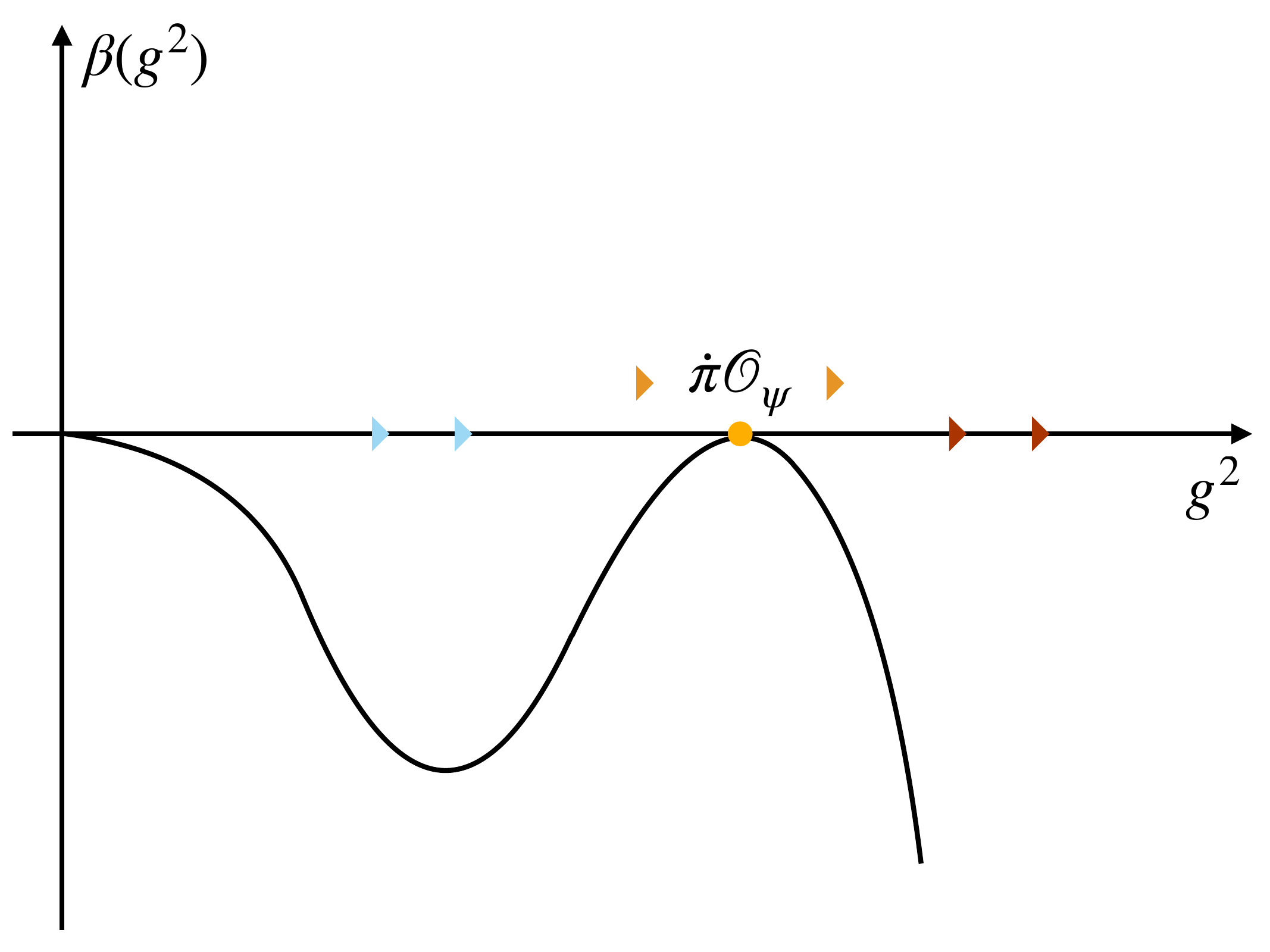}
    \caption{The $\beta$-function with merged fixed points}
    \label{fig:beta_function}
\end{figure}

It is long known that there exists a non-trivial IRFP for $\text{SU}(N_c)$ theories with a certain range of $N_f$, which is called the conformal window \cite{Banks:1981nn}. 
The $\beta$-function of $\text{SU}(N_c)$ theories in the asymptotically free phase is given in \cite{Gross:1973id,Caswell:1974gg}
\begin{align}
    \beta(g) = - \beta_0 \frac{g^3}{(4\pi)^2} + \beta_1 \frac{g^5}{(4\pi)^4} + O(g^7), 
    \label{def:beta_function}
\end{align}
where $\beta_0$ and $\beta_1$ are 
\begin{align}
    \beta_0 &= \frac{11}{3} N_c - \frac{2}{3} N_f, \\
    \beta_1 &= -\frac{34}{3} N_c^2 + \frac{10}{3} N_c N_f + \frac{N_c^2-1}{N_c} N_f. 
\end{align}
The theory could arrive at an IRFP if the first two terms have opposite signs. 
Up to two loops, the coupling constant $g_c^2$ will be 
\begin{align}
    &\frac{g_c^2}{(4\pi)^2} = \frac{\beta_0}{\beta_1} = \frac{11N_c^2-2N_c N_f}{13N_c^2 N_f-3N_f-34N_c^3}. 
    \label{def:a_2_loops}
\end{align}
From the $\beta$-function we notice that, generally in the perturbative regime $g\lesssim4\pi$, the IRFP is stable \cite{Banks:1981nn}, which means that a small relevant deformation will not drag the Dark Walker into the gapped phase. 
Therefore, we have to look for special gauge theories. 
Fortunately, recent studies in lattice QCD (e.g. \cite{Butt:2024kxi,Witzel:2024bly}) have revealed that, for the $\text{SU(2)}$ theory with four flavors of massless Dirac fermions and the $\text{SU}(3)$ theory with eight flavors of massless Dirac fermions, the ultraviolet (UV) and IR fixed points will merge, as illustrated in Figure \ref{fig:beta_function}. 

In the following, we will take the Dark Walker to be an $\text{SU}(3)_D$ theory with $N_f=8$ flavors of massless Dirac fermions. 
This is an illustrator of our \textit{Dark Walker mechanism}. 
In general, this scenario should hold for all gauge models with a merged fixed point (MFP). 
This kind of mass generation is called the \textit{symmetric mass generation} (SMG), a phenomenon where the fermions acquire mass gaps while the chiral symmetry remains unbroken. 
For a recent review, the readers can refer to \cite{Wang:2022ucy}. 
\newline

\noindent{\bf During Inflation.}---In this work, we choose the value of $H_{\text{inf}}$ to be the current upper bound value from \cite{Planck:2018jri}, which is $H_{\text{inf}}\sim2.5\times10^{-5} M_{\text{pl}} \sim 6.0\times10^{13}\text{ GeV}$. 
In the context of cosmology, it is convenient to use the number of $e$-folds $N$ to measure the ratio between scales. 
There are $\sim10$ $e$-folds between the Planck mass scale and the Hubble scale during inflation. 
Once we assume the inflation starts from the $M_{\text{pl}}$, the working energy scale during inflation flows from $M_{\text{pl}}$ to $H_{\text{inf}}$, and stays at $H_{\text{inf}}$ until the end of inflation. 

If we start from a weak gauge coupling in the UV, we can estimate the number of $e$-folds needed to arrive near the MFP using 
\begin{align}
    &\frac{\dint g}{\dint N} = \beta_0\frac{g^3}{(4\pi)^2} - \beta_1 \frac{g^5}{(4\pi)^4} \nonumber \\
    &\begin{aligned}
    \Rightarrow\ N_{\text{cw}} &= \frac{8\pi^2}{\beta_0}\left(\frac{1}{g_0^2}-\frac{1}{g_c^2}\right) \\
    &+ \frac{\beta_1}{2\beta_0^2}\log\left(\frac{g_c^2}{g_0^2}\frac{16\pi^2\beta_0-g_0^2\beta_1}{16\pi^2\beta_0-g_c^2\beta_1}\right). 
    \end{aligned}
    \label{eqn:beta_inflation}
\end{align}
We will take the critical gauge coupling $g_c^2$ to be the lattice result of \cite{Witzel:2024bly}. 
The critical \textit{gradient-flow} coupling $g_{\text{GF}}^2\approx25$ will give us $g_c^2\lesssim12$, with reference to \cite{Luscher:2010iy}. 
Surely, this one-loop matching in \cite{Luscher:2010iy} will break up for strong couplings, but all we need is an upper bound for the number of $e$-folds needed. 
We will restrict the ranges of $N_{\text{cw}}$ and the initial value of the gauge coupling $g_0$ after constructing the effective interactions. 

Once our model reaches the MFP, the gauge theory \eqref{def:UV_gauge_theory} becomes a conformal theory, and we cannot use this Lagrangian. 
Generally, it is not easy to derive the operator spectrum and the operator product expansion (OPE) coefficients between operators from the UV theory \eqref{def:UV_gauge_theory}. 
Nevertheless, we can focus on a few operators that we expect to see from the UV theory. 
An important one is the fermion bilinear operator $\mathcal{O}_\psi\equiv\bar{\psi}\psi$, as the anomalous dimension $\gamma_m$ of $\mathcal{O}_\psi$ will be greater than 1 near the MFP, studied in \cite{Cheng:2013eu}. 
This property will trigger ``walking'' and drag the Dark Walker into the gapped phase, as we will see below. 
Another possible primary scalar operator is $\mathcal{O}_{F}\equiv\text{Tr}F^2$, where the scaling dimension $\Delta_F\lesssim4$. 
The explicit value can only be found from lattice simulations. 
The operator $\mathcal{O}_\psi$ has the scaling dimension $\Delta_\psi=3-\gamma_m\approx 1.8$ from \cite{Cheng:2013eu}, and we assume $\Delta_{F}\approx3.9$ for $\mathcal{O}_F$. 

Using the language of effective field theory (EFT) of inflation \cite{Cheung:2007st,Lee:2016vti}, we can construct the linear interactions between the scalar primary operators $\mathcal{O}_\Delta$ and the \textit{scalar perturbation} $\pi$ as given in \cite{Green:2013rd,Pimentel:2025rds}: 
\begin{align}
    &\mathcal{L}^{(s=0)}_{\pi\mathcal{O}} \equiv \frac{1}{2}\mu^{2-\Delta} M_{\text{pl}}|{\dot{H}_{\text{inf}}}|^{1/2} \mathcal{O}_\Delta (g^{00}+1) \nonumber \\
    &= \frac{1}{2}\mu^{2-\Delta} M_{\text{pl}}|{\dot{H}_{\text{inf}}}|^{1/2} \left(-2\dot{\pi}\mathcal{O}_\Delta + (\partial_\mu\pi)^2\mathcal{O}_\Delta\right). 
     \label{def:EFT_inflation}
\end{align}
It is worth mentioning that now we should regard the unparticles $\mathcal{O}_\psi$ and $\mathcal{O}_F$ as composite particles, which means, for example, $\dot{\pi}\mathcal{O}_\psi$ is a linear mixing instead of a Yukawa coupling. 
The dimension of the linear mixing $\dot{\pi}\mathcal{O}_\psi$ is less than four, making this linear mixing relevant. 
This interaction will walk the theory from the MFP into the gapped phase, while a small relevant interaction will not affect the overall dynamics of inflation. 

Since now the power of the coupling constant depends on $\Delta_\psi$ and $\Delta_F$, we need the effective coupling constant $\mu_\psi$ to be smaller than $H_{\text{inf}}$ and $\mu_F$ to be larger than $H_{\text{inf}}$, in order to have a sufficient perturbative description. 
In a slow-roll inflation model, it is natural to take $\mu^{2-\Delta}=\dot{\phi}/\Lambda_{\text{cw}}^{\Delta}$, where $\Lambda_{\text{cw}}$ is the energy scale to form the unparticles. 
Since $g_c/4\pi\lesssim1$, \eqref{eqn:beta_inflation} is enough for an estimate. 
In order to have a sufficient separation of scales to perform perturbative analysis, we require the lower bound of $\Lambda_{\text{cw}}$ to be $\Lambda_{\text{cw}}^\Delta\gtrsim\dot{\phi}H_{\text{inf}}^{\Delta-2}$ for all $\Delta$'s we consider. 
Regarding all the requirements above, the number of $e$-folds needed to arrive near the MFP will be restricted to $0\lesssim N_{\text{cw}}\lesssim5.8$, and the initial value $g_0$ will be limited to $1.7\lesssim g_0\lesssim3.4$. 
Explicitly, 
\begin{align}
    7.3\times10^{15}\text{ GeV}\lesssim\Lambda_{\text{cw}}\lesssim M_{\text{pl}}. 
\end{align}
In addition, assuming that $\langle\mathcal{O}_\Delta^3\rangle$ is negligible, the bispectrum is determined by the quadratic derivative coupling $(\partial_\mu\pi)^2\mathcal{O}_\Delta$, as discussed in \cite{Pimentel:2025rds}. 
Having established the EFT of inflation with unparticles, we now turn to its cosmological signatures. 
\newline

\begin{figure}[h]
    \centering
    \includegraphics[width=1.\linewidth]{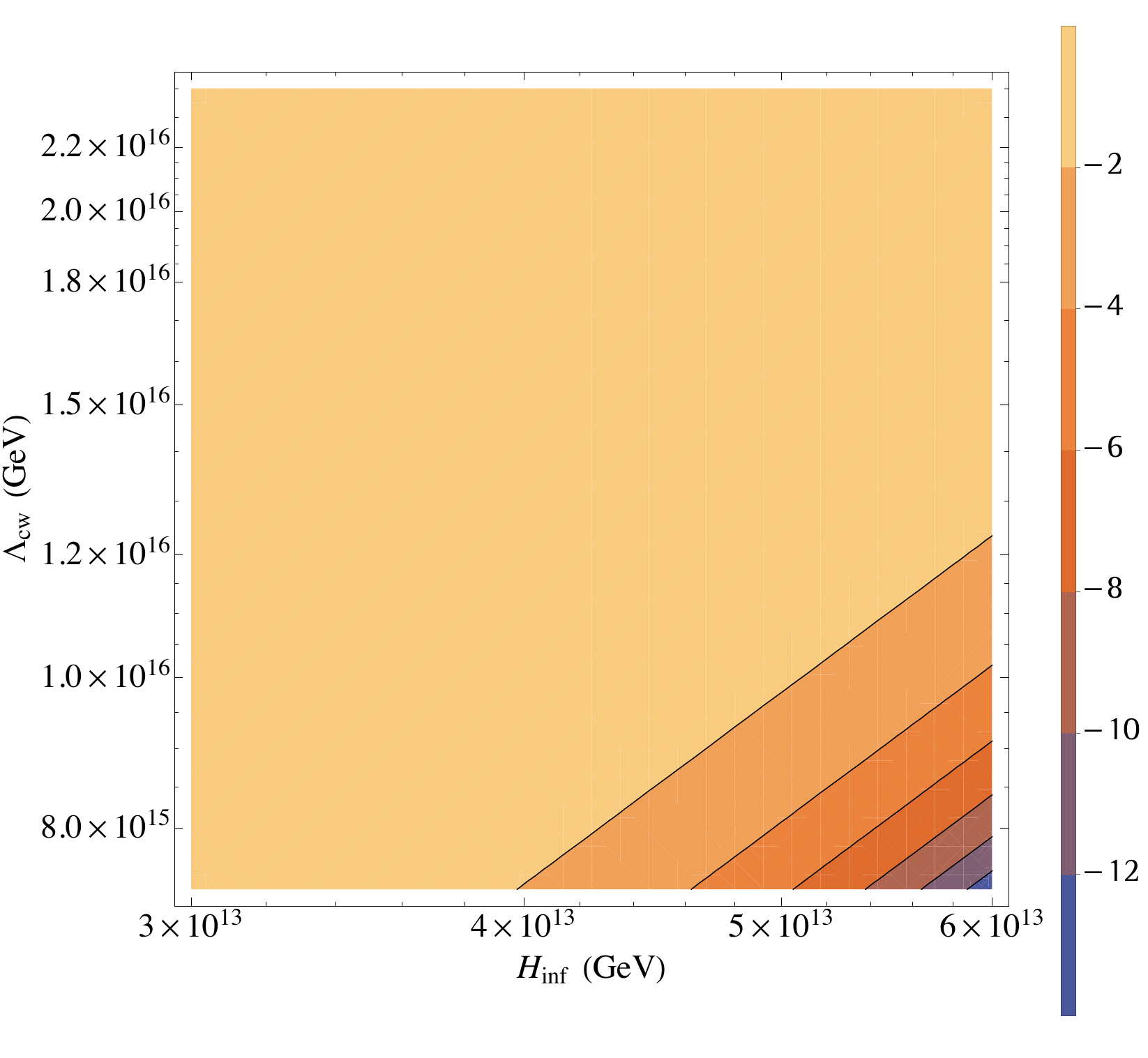}
    \caption{The contour plot of $f_{\text{NL},\psi}$ at $O(1)$ for a range of $H_{\text{inf}}$ and $\Lambda_{\text{cw}}$. The magnitude of $f_{\text{NL}}$ is sensitive to the value of $H_{\text{inf}}$. }
    \label{fig:fNL}
\end{figure}

\noindent{\it Non-Gaussianity.}---The cosmological bispectrum can be calculated using the results in \cite{Pimentel:2025rds}. 
The non-Gaussianity $f_{\text{NL}}$ is defined as 
\begin{align}
    f_{\text{NL}} \equiv \frac{5}{18} \frac{(k_1k_2k_3)^2}{(2\pi^2)^2\Delta_{\zeta}^4(k)} \langle\zeta_{\vec{k}_1}\zeta_{\vec{k}_2}\zeta_{\vec{k}_3}\rangle'\big|_{k_1=k_2=k_3=k}, 
\end{align}
where 
\begin{align}
    \langle\zeta_{\vec{k}_1}\zeta_{\vec{k}_2}\zeta_{\vec{k}_3}\rangle \equiv (2\pi)^3\delta^{(3)}(\vec{k}_1+\vec{k}_2+\vec{k}_3) \langle\zeta_{\vec{k}_1}\zeta_{\vec{k}_2}\zeta_{\vec{k}_3}\rangle' \nonumber
\end{align}
is the three-point correlation function of \textit{curvature perturbation} $\zeta$. 
$\Delta_\zeta^2 = 2.1\times10^{-9}$ is the dimensionless \textit{power spectrum} of $\zeta$ from \cite{Planck:2018vyg}. 
At linear order, the curvature perturbation $\zeta$ relates to the scalar perturbation $\pi$ via \begin{align}
    \zeta = -H_{\text{inf}}\ \pi. 
\end{align}
Taking the slow-roll parameter $\epsilon$ to be $0.0097$ according to \cite{Planck:2018jri}, we find that $-1.2\times10^{-8}\lesssim f_{\text{NL},\psi} \lesssim -13$ and $|f_{\text{NL},F}|<10^{-10}$. 
The full expression of $f_{\text{NL}}$ is presented in Appendix \ref{app:fNL}. 
The $f_{\text{NL}}$ with $\mathcal{O}_\psi$ is desirable because it has the potential to be $O(1)$, which is around the center of the latest observation $f_{\text{NL}}=6\pm46$ \cite{Jung:2025nss}, while large enough to be detectable with the experiments in the near future. 
The $f_{\text{NL},F}$ is too small to be detected. 
Even the futuristic 21-cm tomography can only give a limiting strength of non-Gaussianity for observability about $f_{\text{NL}}\gtrsim10^{-4}$ \cite{Meerburg:2016zdz}. 
Notice that the choice of $H_{\text{inf}}$ will significantly affect the value of $f_{\text{NL}}$, as indicated in Figure \ref{fig:fNL}. 

\begin{figure}[h]
    \centering
    \includegraphics[width=1.\linewidth]{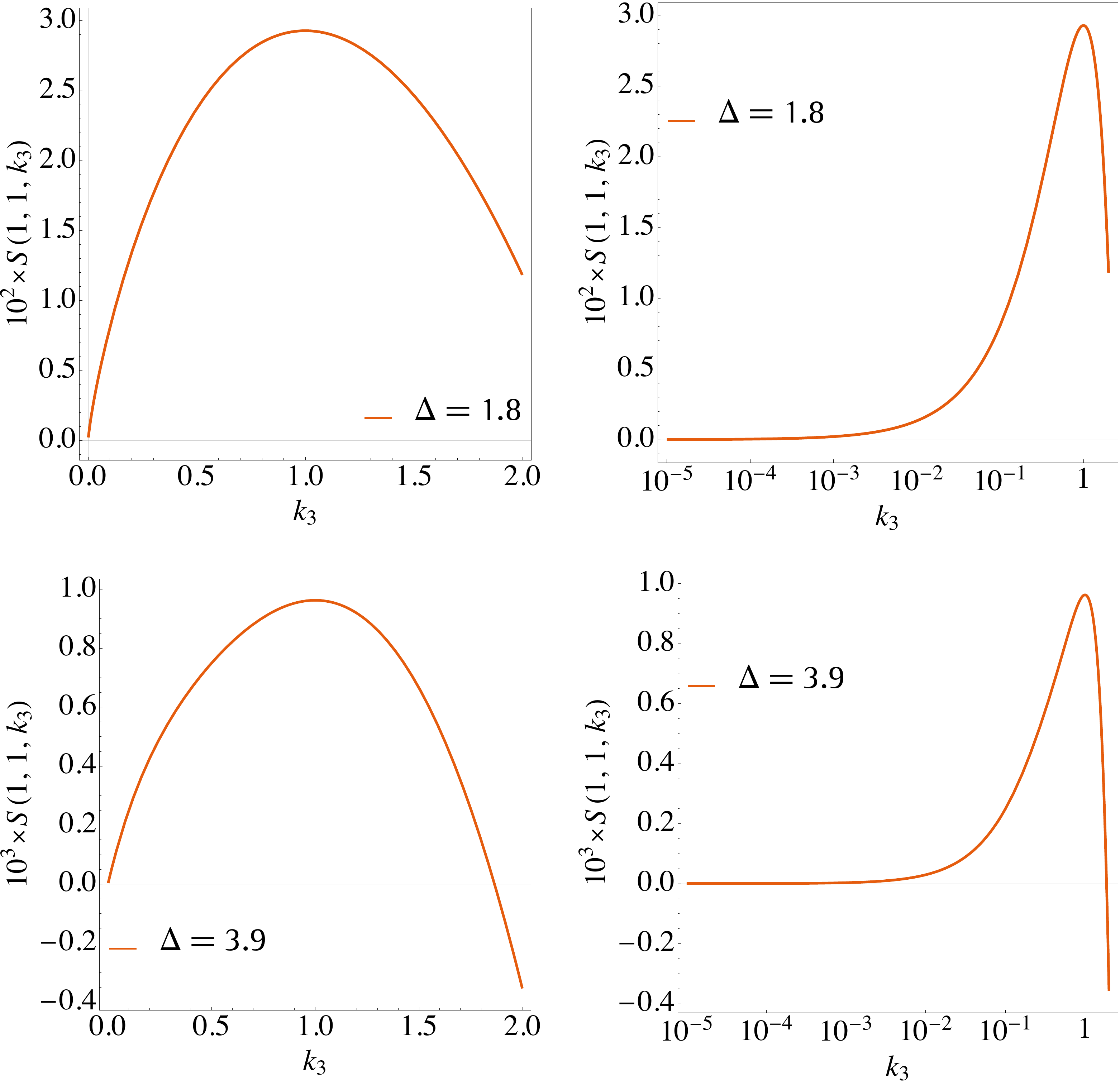}
    \caption{Shape functions for $\Delta=1.8$ and $\Delta=3.9$. We rescale the magnitudes by $10^2$ for $\Delta=1.8$ and $10^3$ for $\Delta=3.9$ to set the plot value within $O(1)$. }
    \label{fig:shape_functions}
\end{figure}

Figure \ref{fig:shape_functions} shows the shape functions with $\mathcal{O}_\psi$ and $\mathcal{O}_F$ exchanges. 
Both of them are near-equilateral, while the squeezed limit behaviors degenerate with the equilateral non-Gaussianity. 
Only the full shapes of the bispectra can break this degeneracy, as emphasized in \cite{Pimentel:2025rds}. 
\newline

\noindent{\bf After inflation.}---This section connects the inflationary phase to the present-day dark matter abundance. 
After the inflation ends, the universe becomes almost flat. 
The interactions $\dot{\pi}\mathcal{O}_\psi$ and $(\partial_\mu\pi)^2\mathcal{O}_\psi$ will freeze, since the energy scale becomes too small compared to the EFT of inflation scale $M_{\text{pl}}|{\dot{H}_{\text{inf}}}|^{1/2}$. 
However, even though the linear mixing interaction $\dot{\pi}\mathcal{O}_\psi$ remains small before it freezes, its nature as a relevant operator will drag our model into the gapped phase. 
The gapped states are natural DM candidates. 
The masses of the lowest lying states of pseudoscalars (PSs) and vector mesons (Vs) can be read from the lattice QCD results \cite{Witzel:2024bly,Rinaldi:2015axa}, where the mass of the lightest vector mesons $m_{\text{V}}$ is 2.2 times the mass of the lightest pseudoscalars $m_{\text{PS}}$. 
We will assume that the DM is made of the lightest pseudoscalars. 
We will use the \textit{freeze-in} process to produce DM, mainly following the framework discussed in \cite{Redi:2021ipn,Hong:2019nwd}. 
It is worth mentioning that if the fermions have light masses in the UV \cite{LatKMI:2016xxi,LatticeStrongDynamics:2023bqp}, the DM could become the dilaton forbidden dark matter, as discussed in \cite{Appelquist:2024koa}. 

If we assume that most of the energy from the inflaton is transferred to the SM after inflation, the SM is thermal at the reheating temperature $T_{R}$. 
Therefore, energy can be extracted from it and injected into the DS via a freeze-in process, due to the coupling between the DS operators and Higgs.\footnote{We thank the anonymous reviewer for bringing this up.}
The model can be described by 
\begin{align}
    \mathcal{L}_{\text{SM}} + \mathcal{L}_{\text{CFT}} + \sum \lambda\frac{\mathcal{O}_{\text{SM}} \mathcal{O}_{\Delta}}{\Lambda^{d}}, \  d=d_{\text{SM}} + \Delta - 4. 
    \label{def:DM_SM_coupling}
\end{align}
Here $\mathcal{O}_{\text{SM}}$ is an SM singlet operator with dimension $d_{\text{SM}}$ and we will work in the regime of small dimensionless Wilson coefficient $\lambda\ll1$. 
In the Higgs portal, the choice of scalar operators $\mathcal{O}_{\text{SM}}$ in the UV up to dimension four can be $|\text{H}|^2$ or $|D_\mu\text{H}|^2$. 
From \cite{Hong:2019nwd} and also later, when $d>1/2$, the freeze-in processes are UV-dominated, and when $d<1/2$, the processes are IR-dominated. 
We assume the reheating temperature $T_R$ to be the highest temperature of the thermal bath, and it is much higher than the electroweak (EW) symmetry breaking scale $\sim100\text{ GeV}$. 
The UV-dominating interaction will be $|D_\mu\text{H}|^2\mathcal{O}_\psi$, whose $d=1.8$. 
A natural choice for the effective coupling is $\Lambda^{\Delta}\equiv\Lambda_{\text{cw}}^\Delta$. 
The interaction $|D_\mu\text{H}|^2\mathcal{O}_F$, and the interactions between currents and stress tensors, e.g. $J_{\text{SM}}J_{\text{DS}}$, $T_{\text{SM}}T_{\text{DS}}$, etc., will be sub-dominant, because of the large scaling dimensions. 
As for the interaction $|\text{H}|^2\mathcal{O}_F$, the effective coupling is natural to be $\Lambda^{\Delta-2}\equiv\Lambda_{\text{cw}}^\Delta/\langle v\rangle^2$, where $\langle v\rangle$ is the vacuum expectation value (VEV) of the Higgs. 
This coupling is much suppressed in the UV regime. 
Nonetheless, the indirect interaction between the Higgs and $\mathcal{O}_F$ will also be sub-dominate and it is natural not to include this interaction in the UV freeze-in processes. 

In the IR-dominant regime, we expand Higgs near the VEV as $\text{H}=(v+h)/\sqrt{2}$ in \eqref{def:DM_SM_coupling}. 
The dominant effective interaction will become 
\begin{align}
    \mathcal{L}_{\text{int}} = \lambda\frac{m^2}{\Lambda_{\text{cw}}^\Delta} \frac{v}{\sqrt{2}} h \mathcal{O}_{\text{CFT}}, 
    \label{def:IR_int_DS_SM}
\end{align}
where $m$ is a mass parameter, which we will further assume to be the Higgs mass $m_h\sim125\text{ GeV}$. 
The VEV of Higgs after EW symmetry breaking is $v\sim246\text{ GeV}$. 
We assume that, in our scenario, the UV- and IR-dominant regimes are well-separated, and the energy density in the DS will remain far below the equilibrium with the SM. 
Thus the decay $\text{CFT}\rightarrow\text{SM}$ can be neglected. 
For now we will take $T_R/m_h>O(10)$, and we will see in the following that $T_R$ is actually lower-bounded by the permitted DM mass range. 

We assume that the Dark Walker completely breaks the conformal symmetries and develops a mass gap $m_{\text{PS}}$ at scale $T_D \sim m_{\text{PS}}$. 
The SM can be treated as a free CFT with $m_{\text{PS}}\lesssim100\text{ GeV}$ \cite{Redi:2021ipn}. 
Thus the UV DS production can be treated as the Dark Walker CFT states from the SM CFT. 
From the classification of mass generation mechanisms in \cite{Wang:2022ucy}, we know that in the SMG case, the system is anomaly-free, and potentially the effective interactions between the pseudoscalars will be strongly coupled, since there is no small scale to suppress the interactions. 
The particle number density in the DS will thus not be conserved, and a more relevant quantity is the energy density flowing into the DS. 
Assuming that the DS evolves relativistically, the energy density $\rho_{\text{CFT}}$ of the DS CFT is controlled by the Boltzmann equation: 
\begin{align}
    \dot{\rho}_{\text{CFT}} = - 4H(T)\rho_{\text{CFT}} + \gamma_1(T), 
    \label{def:Boltzmann_eqn}
\end{align}
with the energy injection rate $\gamma_1$ parametrized in the standard way as 
\begin{align}
    \gamma_1(T) = \kappa_1 T^5 \lambda^2 \left(\frac{T}{\Lambda_{\text{cw}}}\right)^{2d}. 
\end{align}

We assume that the initial energy density of DS and the energy loss from the thermal bath are both negligible. 
We can find the energy density to be 
\begin{align}
    \rho(T) = \frac{3\sqrt{10}}{\sqrt{g_*}\pi} \frac{\kappa_1 \lambda^2 T_R^{2d}}{(2d-1)\Lambda_{\text{cw}}^{2d}} \frac{M_{\text{pl}}}{T_R} T^4. 
    \label{eqn:energy_density}
\end{align}
Here $g_*$ is the effective degrees of freedom of the relativistic species, and since the energy density of the DS remains small compared to the energy density of the SM, we will take the value for the SM. 
Notice that we have assumed that this DS production is UV-dominated, which corresponds to $d>1/2$ interactions. 

By considering the $2\rightarrow\text{CFT}$ scattering process $|D_\mu\text{H}|^2\rightarrow\text{CFT}$, \cite{Redi:2021ipn} has computed the energy injection rate coefficient $\kappa_1$ to be 
\begin{align}
    \kappa_1 = \frac{a_{\mathcal{O}_i}a_{\mathcal{O}_f}}{16\pi^5} \frac{\Gamma(d_i+d_f-1)\Gamma(d_i+d_f-3)}{\Gamma(d_i)\Gamma(d_i-1)\Gamma(d_f)\Gamma(d_f-1)}, 
\end{align}
where $a_{\mathcal{O}}$ is the normalization factor of the two-point functions, $i$ stands for the initial state and $f$ denotes the final state. 
For the scattering process $|D_\mu\text{H}|^2\rightarrow\text{CFT}$, we take the normalization factors to be $a_{|D_\mu\text{H}|^2}=48$ and $a_{\text{CFT}}=2\pi^2$. 
Upon collecting all the factors, the coefficient is $\kappa_1=0.44$ for $|D_\mu\text{H}|^2\rightarrow\mathcal{O}_\psi$. 
The energy density from the UV-dominant process will be 
\begin{align}
    \rho_{\text{CFT, UV}} = 1.2\times10^{17}\lambda^2 T_R^{2.6}T^4\text{GeV}/\Lambda_{\text{cw}}^{3.6}. 
    \label{res:rho_UV}
\end{align}

The IR-dominant interaction \eqref{def:IR_int_DS_SM} will give the Higgs decay contribution \cite{Georgi:2007ek,Grinstein:2008qk,Hong:2019nwd} 
\begin{align}
    \gamma'_1(T) &= n_h \langle\Gamma(h\rightarrow\text{CFT})E\rangle \nonumber \\
    &= f_\Delta \frac{\lambda^2 m_h^{2\Delta+2}v^2T}{\Lambda_{\text{cw}}^{2\Delta}}K_2\left(\frac{m_h}{T}\right), 
\end{align}
where $f_\Delta\equiv4^{-\Delta}\pi^{1/2-2\Delta}\Gamma(\Delta+1/2)/(\Gamma(\Delta-1)\Gamma(2\Delta))$ and $K_\nu(x)$ is the modified Bessel functions of the second kind. 
The process roughly starts around the EW scale $\sim v$ and ends when the SM temperature reaches the Higgs mass $m_h$ \cite{Hong:2019nwd}. 
After the temperature reaches the Higgs mass, the Higgs decouples from the thermal bath and the decay will be exponentially suppressed. 
We can take the limit $m_h\rightarrow0$ in the thermal average computation and simplify the decay contribution: 
\begin{align}
    n_h \langle\Gamma(h\rightarrow\text{CFT})E\rangle = 2f_\Delta \frac{\lambda^2 m_h^{2\Delta}}{\Lambda_{\text{cw}}^{2\Delta}}v^2T^3. 
\end{align}
Solving the Boltzmann equation \eqref{def:Boltzmann_eqn} and taking the temperature to be $m_h$ will give us 
\begin{align}
    \rho_{\text{CFT, IR}} (m_h) &= \frac{\sqrt{5}M_{\text{pl}}f_\Delta\lambda^2m_h^{2\Delta+4}}{\pi^{3/2}\sqrt{g_*}v\Lambda_{\text{cw}}^{2\Delta}} \left(\frac{v^3}{m_h^3}-1\right) \nonumber \\
    &= 3.9\times10^{20}\lambda^2 m_h^{3.6}/\Lambda_{\text{cw}}^{3.6}\text{ GeV}^4. 
\end{align}
When $T\sim m_h$, the ratio between $\rho_{\text{CFT, UV}}$ and $\rho_{\text{CFT, IR}}$ is 
\begin{align}
    \frac{\rho_{\text{CFT, UV}}}{\rho_{\text{CFT, IR}}} = 2.1 \times 10^{-3} \left(\frac{T_R}{\text{GeV}}\right)^{2.6} \gtrsim 2.4\times10^5. 
\end{align}
Thus the IR-dominant contribution is sub-dominant, and we will neglect it when estimating the DM masses. 

Since the DS is strongly interacting, the DS will self-thermalize. 
Assuming that the thermalization is instantaneous and neglecting the loss of energy, the DS temperature at thermal equilibrium $T_D$ relates to the energy density as 
\begin{align}
    \rho = \frac{\pi^2}{30} g_D T_D^4, 
    \label{eqn:energy_density_DS_equi}
\end{align}
where $g_D$ is the number of degrees of freedom in the DS before the phase transition. 
We will see at the end of this section that this is a self-consistent assumption. 
The UV effective degrees of freedom of \eqref{def:UV_gauge_theory} is $100$. 
Flowing to the near-IRFP will require corrections to the effective degrees of freedom, which is beyond the scope of this work. 
We will still take $g_D \sim O(10^2)$. 
Combining \eqref{res:rho_UV} and \eqref{eqn:energy_density_DS_equi}, we can define the ratio between the DS temperature and the SM temperature to be 
\begin{align}
    &\xi \equiv \frac{T_D}{T} = 2.5\times10^{4} \times \left(\frac{\lambda^2T_R^{2.6}\text{GeV}}{g_D\Lambda_{\text{cw}}^{3.6}}\right)^{1/4}. 
    \label{def:ratio_T_D_and_T}
\end{align}
From here the DS starts to have a different temperature, but we assume the energy density to remain small so it will not affect the IR processes. 
We assume that the DM particles are non-relativistic and decouple from the SM, while all injected energy becomes DM masses. 
The DM energy density today is $\rho_{\text{DM},0} = \rho_{\text{CFT}}(T_c) s_{0}/s_{\text{SM}}(T_c/\xi)$, where the SM entropy density is $s=(2\pi^2/45)g_*T^3$. 
The DM relic abundance is defined as 
\begin{align}
    &\frac{\Omega_{\text{DM}}h^2}{0.12} = \frac{\rho_{\text{DM},0}}{\rho_{\text{crit},0}} \frac{h^2}{0.12} = \frac{\rho_{\text{CFT}}(T_c)}{0.36\text{ eV} \times s_{\text{SM}}(T_c/\xi)} \nonumber \\
    =& 3.5\times10^{36}\frac{T_c}{\text{GeV}} \left(\frac{\lambda^2T_R^{2.6}\text{GeV}}{\Lambda_{\text{cw}}^{3.6}}\right)^{7/4}. 
    \label{def:relic_abundance}
\end{align}

\begin{figure}[h]
    \centering
    \includegraphics[width=1.\linewidth]{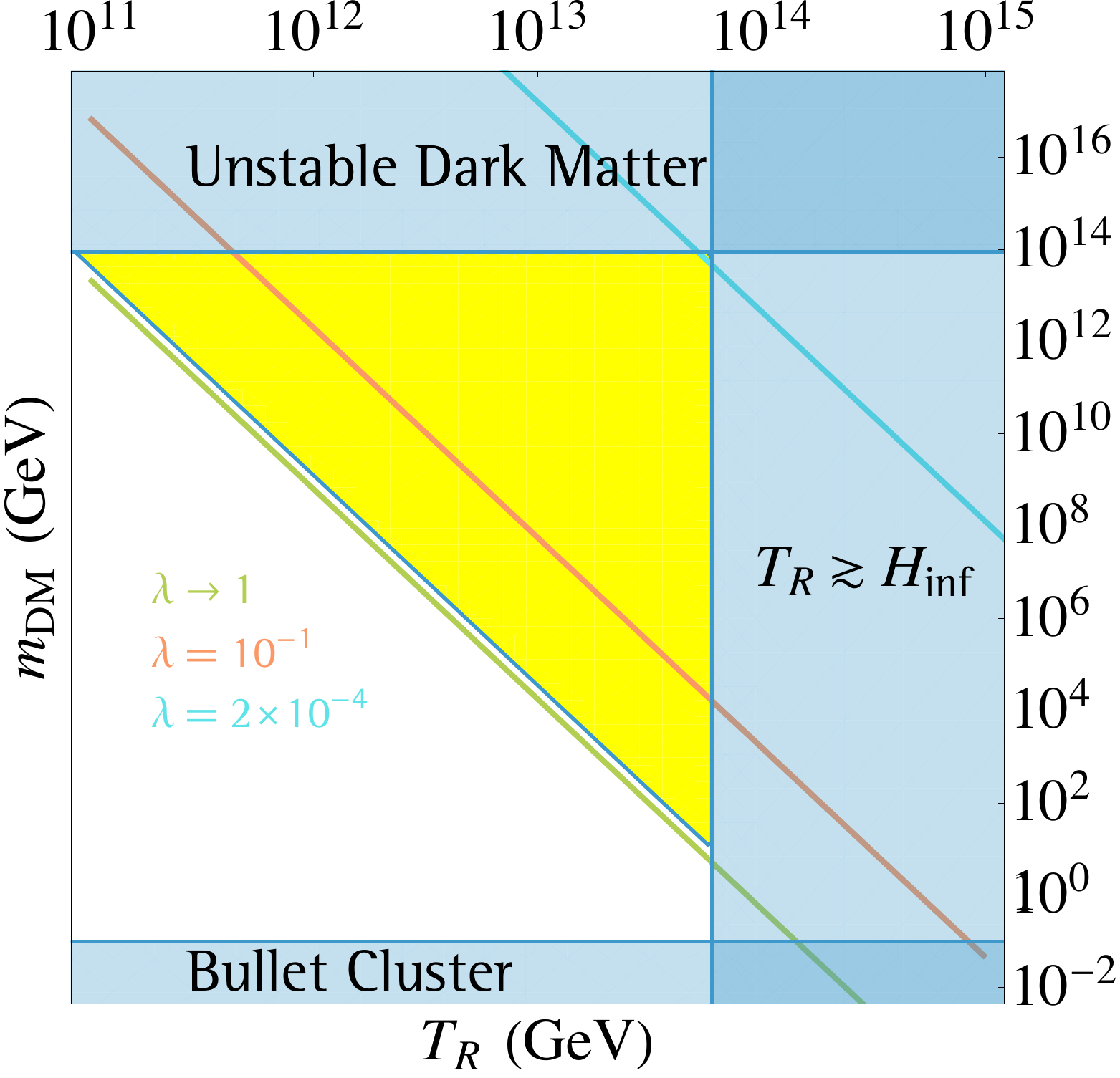}
    \caption{The relation between the DM particle mass and the Wilson coefficient $\lambda$. The yellow region is the permitted mass range with respect to $T_R$. We take $\Lambda_{\text{cw}}$ to be the lowest value, in order to find the widest mass range. The blue regions are excluded by the Bullet Cluster observations, inflationary Hubble constant and DM lifetime. }
    \label{fig:DM_mass_relation}
\end{figure}

Assuming the confining temperature to be $T_c\sim m_{\text{PS}}$, the DM particle mass will be 
\begin{align}
    &\frac{\Omega_{\text{DM}}h^2}{0.12} \approx 1 \nonumber \\
    \Rightarrow& m_{\text{PS}} \sim 2.9\times10^{-37} \left(\frac{\Lambda_{\text{cw}}^{3.6}}{\lambda^2T_R^{2.6}\text{GeV}}\right)^{7/4}\text{ GeV}. 
\end{align}
We can see that the mass of DM particles increases with larger $\Lambda_{\text{cw}}$ and smaller $T_R$. 
If we take the lowest value of $\Lambda_{\text{cw}}$, the DM mass becomes 
\begin{align}
    m_{\text{PS}} \sim 2.5\times10^{63} \left(\frac{\text{GeV}^{2.6}}{\lambda^2T_R^{2.6}}\right)^{7/4}\text{ GeV}. 
\end{align}
By estimating the DM lifetime $\tau_{\text{DM}} \sim M_{\text{pl}}^4/m_{\text{PS}}^5>10^{28}\text{ s}$, we can setup the upper bound for the DM mass once we consider the decay into gravitons and $\gamma$-rays \cite{Redi:2021ipn,Blanco:2018esa}. 
The decay into Higgs and other SM particles will be much suppressed by the small portal coupling, as for typical freeze-in processes, and thus will give a looser bound. 
A lower bound can be established by considering standard DM self-interaction, taking $\langle\sigma v\rangle\sim\pi/m_{\text{PS}}^2$. 
The region $m_{\text{PS}}\lesssim10^{-1}\text{ GeV}$ will be excluded from the Bullet Cluster observations \cite{Redi:2021ipn}. 
Assuming that the reheating temperature cannot exceed the inflationary Hubble constant and taking the upper and lower bounds of DM mass, we can plot the region of permitted masses, as shown in Figure \ref{fig:DM_mass_relation}. 

Now we look back and argue for the self-consistency of DS self-thermalization. 
From the plot we can see that $\lambda\gtrsim2\times10^{-4}$, for the lowest value of $\Lambda_{\text{cw}}$. 
We assume that the self-thermalization of DS is dominated by $2\rightarrow2$ processes. 
Roughly we can take the interaction rate $\Gamma\sim n_D\langle\sigma v\rangle$. 
The number density of DS is $n_D=\zeta(3)g_DT_D^3/\pi^2\approx12T_D^3$. 
Taking the cross section $\langle\sigma v\rangle$ to be $\langle\sigma v\rangle\sim\alpha^2/T_D^2$, where $\alpha$ is an effective coupling, the interaction rate becomes $\Gamma=12\alpha^2T_D$. 
We have 
\begin{align}
    &\frac{\Gamma}{H} \sim 9.1\times10^5\frac{\alpha^2M_{\text{pl}}}{\sqrt{g_*}T}\left(\frac{\lambda^2T_R^{2.6}\text{GeV}}{g_D\Lambda_{\text{cw}}^{3.6}}\right)^{1/4} \\
    &\sim 9.1\times10^5\frac{\alpha^2M_{\text{pl}}}{\sqrt{g_*}g_D^{1/4}T} \left(3.5\times10^{36}\frac{m_{\text{PS}}}{\text{GeV}} \right)^{-1/7}
\end{align}
Since we assume the self-thermalization to happen instantaneously, we can take $T$ of SM to be $T_R$. 
The ratio $\Gamma/H$ will become 
\begin{align}
    \frac{\Gamma}{H} = \frac{4.0\times10^{17}\alpha^2\text{ GeV}}{\left(m_{\text{PS}}/\text{GeV}\right)^{1/7}T_R}. 
\end{align}
Thus, for a natural effective coupling $\alpha\sim O(1)$, the ratio $\Gamma/H\gtrsim O(10^2)$, so the self-thermalization is sufficient. 
From Figure \ref{fig:DM_mass_relation}, we can also find that the IR-dominant process is not only sub-dominant but also kinetically forbidden. 
\newline

\noindent{\bf Discussion.}---A simple gauge theory at the lower edge of the conformal window in the DS opens up new possibilities for beyond-Standard-Model (BSM) physics and DM candidates. 
The Dark Walker model has the potential to produce both the primordial non-Gaussianity, which is large enough to have the possibility of being detectable, and the observed relic abundance of dark matter through freeze-in. 
As expected, several parameters must still be fixed by hand, though their values remain within natural ranges. 
The non-Gaussianity signal is sensitive to the inflationary Hubble scale, the initial gauge coupling---which determines the starting energy scale of the merged fixed point---and the precise anomalous dimensions of scalar primary operators in the conformal phase. 
The dark matter abundance is sensitive to the critical temperature of confinement, the reheating temperature and also the energy scale of unparticle formation. 
These parameters may be constrained more accurately in the future through improved observations of cosmological parameters and lattice QCD studies of the spectrum in the $\text{SU}(3)$ theory with eight flavors. 
With such inputs, more precise predictions of the primordial non-Gaussianity and relic abundance can be achieved. 
This framework offers a concrete and coherent testing ground for the role of strongly coupled sectors in cosmology. 
\newline

\noindent{\bf Acknowledgements}

I express my thanks to Riccardo Barbieri, who encouraged me very helpfully while this work was in progress.
Thanks to Marco Costa, Raffaele-Tito D'Agnolo, Janet Hung, Guilherme Leite Pimentel, Alessandro Strumia and the Reviewers for useful discussions and comments. 

CY is supported by Scuola Normale and by INFN (IS GSS-Pi). 
The research of CY is moreover supported by the ERC (NOTIMEFORCOSMO, 101126304). 
Views and opinions expressed are, however, those of the author(s) only and do not necessarily reflect those of the European Union or the European Research Council Executive Agency. 
Neither the European Union nor the granting authority can be held responsible for them. 

\appendix
\section{Calculation of non-Gaussianity}
\label{app:fNL}
In the main text, we used the EFT language of inflation to express the non-Gaussianity $f_{\text{NL}}$ as 
\begin{align}
    &f_{\text{NL}} \equiv \frac{5}{18} \frac{(k_1k_2k_3)^2}{(2\pi^2)^2\Delta_{\zeta}^4(k)} \langle\zeta_{\vec{k}_1}\zeta_{\vec{k}_2}\zeta_{\vec{k}_3}\rangle'\big|_{k_1=k_2=k_3=k} \nonumber \\
    &= \frac{5}{18} \frac{k^6}{(2\pi^2)^2\Delta_{\zeta}^4(k)} (-H_{\text{inf}})^3 \langle\pi_{\vec{k}_1}\pi_{\vec{k}_2}\pi_{\vec{k}_3}\rangle'\big|_{k_i=k}. 
\end{align}
When the scalar perturbation $\pi$ is canonically normalized as 
\begin{align}
    \pi \equiv \frac{\pi_c}{\sqrt{2}M_{\text{pl}}|{\dot{H}_{\text{inf}}}|^{1/2}} = \frac{\pi_c}{\dot{\phi}}, 
\end{align}
the $\pi_c$ will behave exactly the same as $\delta\phi$. 
Combining with the results in \cite{Pimentel:2025rds}, $f_{\text{NL}}$ can be further written as 
\begin{align}
    f_{\text{NL}} =& \frac{5}{18} \frac{2^2 k^6}{(2\pi^2)^2\Delta_{\zeta}^4(k)} \left(-\frac{H_{\text{inf}}}{\dot{\phi}}\right)^3 \left(\frac{\mu}{H_{\text{inf}}}\right)^{4-2\Delta} \nonumber \\
    \times& \frac{H_{\text{inf}}^3}{(2k^3)^3} \mathcal{B}_\phi\left(k,k,k\right). 
\end{align}
The \textit{bispectrum} $\mathcal{B}_\phi\left(k_1,k_2,k_3\right)$ in our model is 
\begin{align}
    &\mathcal{B}_\phi\left(k_1,k_2,k_3\right) \equiv \frac{\epsilon}{2} k_3^3 U_{12}b(u) + \text{permutations}, \nonumber \\
    &U_{12} (\cdot) \equiv \frac{1}{2} \left(1 - \frac{k_1 k_2}{k_{12}}\partial_{k_{12}}\right) \left(\frac{1-u^2}{u^2}\partial_u (u \cdot)\right), \nonumber \\
    &b(u) = \frac{2u}{u+1} \frac{1}{2-\Delta} {}_2F_1\left(1,2-\Delta;3-\Delta;\frac{2u}{u+1}\right) \nonumber \\
    &- \left(\frac{2u}{u+1}\right)^{2(\Delta-1)} \frac{1}{\Delta-1} {}_2F_1\left(1,\Delta-1;\Delta;\frac{2u}{u+1}\right) \nonumber \\
    &+ \frac{2}{\Delta-1} \frac{2u}{u+1} {}_2F_1\left(1,1;\Delta;\frac{1-u}{1+u}\right), 
\end{align}
where ${}_2F_1$ is the hypergeometric function.

\bibliography{refs}

\end{document}